\documentstyle[twocolumn,aps,epsfig,graphicx]{revtex}
\begin{document}
\author{C. Figueira de Morisson Faria and I. Rotter}
\address{Max Planck Institut f\"{u}r Physik komplexer Systeme, \\
N\"{o}thnitzer Str. 38, D-01187 Dresden, Germany }
\title{High-harmonic generation and periodic level crossings:
time profile and control}
\date{\today}
\maketitle

\begin{abstract}
We investigate high-harmonic generation in closed systems, using the
two-level atom as a simplified model. By means of a windowed Fourier transform
of  the time-dependent dipole acceleration, we extract the main contributions
to this process within a cycle of the driving field. We show that the patterns
obtained can be understood by establishing a parallel between the two-level
atom and the three-step model. In both models, high-harmonic generation is a
consequence of a three-step process, which involves either the continuum and
the ground state, or the adiabatic states of the two-level Hamiltonian.
The knowledge of this physical mechanism allows us to manipulate the adiabatic
states, and consequently the harmonic
spectra, by means of a bichromatic driving field. Furthermore, using
scaling laws, we establish sharp criteria for the invariance of the physical
quantities involved. Consequently, our results can be extended to a broader
parameter range, as for instance those characteristic of solid-state systems 
in strong fields.
\end{abstract}

\section{Introduction}

The generation of high-order harmonics of a strong laser field ($I\sim
10^{14}{\rm W/cm}^{2})$ in gaseous samples, where coherent light in the
extreme ultraviolet regime is obtained from infrared input radiation,
originated a breakthrough in nonlinear optics. In these systems, composed by 
atoms or small molecules,
high-harmonic generation (HHG) is a well-understood issue \cite{hhgreview}%
. These highly nonlinear spectra exhibit very particular features: a
frequency region with harmonics of roughly the same intensities, the
``plateau'', and a sharp decrease in the harmonic yield at
the plateau's high-energy
end, the ``cutoff''. Since the early nineties, not only these features have
been investigated, but also the HHG time profile \cite{tprof0,tprof},
physical mechanisms \cite{tla,tstep}, and the propagation of the
harmonic radiation in gaseous media \cite{propag}.\ These studies culminated
with countless proposals of how to control high harmonics, as diverse as for
instance polychromatic \cite{bichro,cfbichro1,cfbichro2} or static \cite{static} fields,
ultrashort pulses \cite{ushort}, or additional potentials \cite{trap}, many
of them having even been realized experimentally \cite{expebic}.

One of the first models proposed to
describe high-harmonic generation in atoms or diatomic molecules was a
two-level atom \cite{tla}. Within this framework, a particularly important
paper is \cite{gauthey}. Therein, it is shown that these harmonics are a
consequence of the population transfer between the field-dependent states
obtained from the diagonalization of the two-level Hamiltonian. This
physical mechanism has not been investigated in detail, and there is a very
simple reason for this apparent lack of interest: it turned out that an at
first sight
completely different physical picture is far more successful in explaining
high-harmonic generation for these systems. This picture, known as ``the
three-step model'', portraits high-harmonic generation as a process in which
an electron leaves an atom at an instant $t_{0}$ (the first step), propagates
in the continuum being accelerated by the field (the second step), and
recombines with the ground state of its parent ion \cite{tstep} at a later
time $t_{1}$, emitting a high-harmonic photon (the third step). This model has
shown that the interplay between a bound state and the continuum, which is not
present in a two-level atom, is essential for a correct physical description of
high-harmonic generation. Thus, the three-step model has
established itself as the paradigm for describing this phenomenon (see,
e.g., \cite{cfmdbound} for a comparison of both models).

Until very recently, only 
gaseous systems were believed to be possible
high-harmonic sources, due to the high intensities involved. However,
nowadays, this picture has changed. With the advent of short
pulses, there are solid-state materials which can survive the necessary
intensity regime, namely  $10^{12}-10^{14}\mathrm{W/cm}^2$ \cite{solid1}. 
 This has led to theoretical studies on
high-harmonic generation in materials such as thin crystals 
\cite{mois98}, or carbon nanotubes \cite{mois2000}. Another example of a
 new and unexpected effect is
for instance carrier-wave Rabi flopping, which has been recently 
measured experimentally \cite{RabFlop}. 

 Furthermore, apart from this entirely new
parameter range, even for considerably lower driving-field intensities, as
for instance $I\sim 10^{6}{\rm W/cm}^{2}$, one may in principle extend the
frequency of far-infrared radiation ($\omega \sim 1{\rm GHz}$) in up to two
orders of magnitude by using adequate materials. For instance, for GaAs/Al$%
_{x}$Ga$_{1-x}$As wells intersubband transitions of $\omega _{0}\sim 1{\rm %
THz}$ may serve this purpose \cite{solid2}.  Apart from these solid-state
 materials, HHG
involving larger molecules is becoming a problem of interest 
\cite{benz1,mois2001}.

For these complex systems, it is not entirely clear whether
bound-to-continuum transitions still yield the most adequate description 
of high-harmonic generation. In
fact, recent studies have shown that, for aromatic molecules, transitions
involving solely bound states are far more important for high-harmonic
generation than the interplay between the ground state and the continuum 
\cite{mois2001}.
Thus, theoretical approaches in which the continuum is not taken into account
may be possibly used to describe this phenomenon in systems as, for instance, 
quantum wells\cite{solid2,reichl,confine,solid3,solid4}. Furthermore,
descriptions of nonlinear optical processes in solids are widely based on
 the Hartree-Fock semiconductor Bloch equations. Under special conditions,
such as low doping density, equal effective masses in both subbands
involved, parallel subbands, and not too wide wells, these equations
 are formally identical to 
those describing the evolution of a two-level atom. 
Otherwise, collective
effects must be taken into account and this analogy is lost
\cite{solid2,confine,solid3,solid4}. 

A common characteristic of all the above-stated systems is their intrincated
internal structure, with the presence, as the external parameters are
varied, of several level crossings. In particular concerning HHG, the
periodic level crossings caused by the temporal dependence of the laser
field are very important \cite{gauthey}. Thus, in order to control the
harmonic spectra also in this context, one needs to understand
the interplay between the population transfer at these crossings and 
high-harmonic generation.

Even in the simplest case for which these level crossings occur, namely a
two-level atom, it is only clear that most of the population transfer
between the field-dressed states takes place at the level crossings.
However, this does not necessarily mean that the population transfers,
within a field cycle, which contribute to the generation of a particular
group of harmonics occur at the level-crossing times. Unanswered questions
in this framework concern not only these times, but also how they depend on
the external-field parameters, such as its intensity and frequency, and how
one can use this information to control the emission spectra of a ``closed",
non-ionizing system. 
Another interesting issue concerns the existence of a one-to-one 
correspondence between the three-step
model and the two-level atom. This was proposed in \cite{gauthey} due to the
different time scales involved in the process, and in \cite{solid2} due to 
a formally identical expression describing population transfers in both
models. In these references, however, there is no proof that this
correspondence really holds.

The answer to these questions is the main objective of this work. The paper
is organized as follows: in Sec. \ref{theory} we briefly discuss the
theoretical background for the studies performed in this paper. 
In the following sections we present our
results. In Sec. \ref{res1}, we concentrate on a detailed analysis of the population transfers
and the time profile of harmonic generation for a monochromatic
field. Subsequently (Sec. \ref{res2}), we provide
concrete examples of how 
an additional driving field may alter the periodic level crossings, and
consequently the harmonic emission
of a closed system. Furthermore, 
we  address the scaling behavior of the physical quantities involved 
(Sec. \ref{res3}), establishing sharp
criteria for their invariance. Finally, in Sec. 
\ref{concl} we close the paper with some concluding remarks.

\section{Background}
\label{theory}

\subsection{Two-level atom}

The simplest case for which level crossings occur, and a widely
used approximation for describing physical systems, is a two-level atom \cite
{AlEber}. Within this picture, the time-dependent wave function is given by

\begin{equation}
\left| \psi (t)\right\rangle =C_{0}(t)\left| \phi _{0}\right\rangle
+C_{1}(t)\left| \phi _{1}\right\rangle ,
\end{equation}
where $C_{n}(t)=\left\langle \phi _{n}\right. \left| \psi (t)\right\rangle $
denotes the overlap of the total wave function with the n-th state of an
arbitrary basis. The evolution of the system is described by the
time-dependent Schr\"{o}dinger equation,  
\begin{equation}
i\frac{d}{dt}\left( 
\begin{array}{c}
C_{0}(t) \\ 
C_{1}(t)
\end{array}
\right) =H\left( 
\begin{array}{c}
C_{0}(t) \\ 
C_{1}(t)
\end{array}
\right) ,
\label{twolev}
\end{equation}
where $H$ is the Hamiltonian matrix, which, in our case, describes an 
atom in an external laser field. We use atomic units throughout. The basis states 
$\left| \phi _{n}\right\rangle $ are
chosen according to the problem at hand. We are particularly
interested in a basis which yields sharp, well-separated level crossings in
the strong-field regime.

A widely used basis are the field-free-states, also known as the ``diabatic
basis''. In this case, the Hamiltonian is given by 
\begin{equation}
H^{D}=\left( 
\begin{array}{ll}
-\omega _{10}/2 & x_{10}E(t) \\ 
x_{10}E(t) & \omega _{10}/2
\end{array}
\right) ,  \label{Hdiab}
\end{equation}
where $\omega _{10}$ is the transition frequency between the field-free
bound states, $E(t)=E_{0}f(t)$ is the external field and $x_{10}$ the dipole
matrix element $\left\langle \phi _{0}^{D}\right| \hat{x}\left| \phi
_{1}^{D}\right\rangle $, where $\left| \phi _{n}^{D}\right\rangle $ denotes
the field-free, ``diabatic'' basis states. This basis is very convenient for
studying level crossings in the low-intensity laser field regime. For strong
laser fields, however, the field-free states are too strongly mixed, such
that a more appropriate basis is needed. Such a basis, which will be called
by us ``exchanged basis'', is obtained applying the unitary transformation

\begin{equation}
U_{D\rightarrow E}=\frac{1}{\sqrt{2}}\left( 
\begin{array}{ll}
1 & 1 \\ 
-1 & 1
\end{array}
\right)  \label{rotation2}
\end{equation}
onto the diabatic basis. The transformation (\ref{rotation2}) was used in 
\cite{gauthey} to interchange the diagonal and the non-diagonal terms of the
Hamiltonian (\ref{Hdiab}). In this case, the exchanged-basis energies
$\varepsilon^E_\pm=\pm x_{10}E(t)$
cross, and the coupling which causes the crossing is effectively given by 
$\omega_{10}/2$. The crossings occur within a time interval
$t_0-t_c<t<t_0+t_c$, where
 $t_c$ is the time for which the off-diagonal and diagonal terms of the Hamiltonian
become equal and $t_0$ is the time for which the field vanishes.
 For strong enough fields, the times over which the 
crossings take place are much
smaller than the period of the driving field. Thus, to first approximation,
one may assume that the crossings take place instantaneously at $t_0$. In the
following we call $t_0$ ``crossing times".

Another important set of basis states are these which diagonalize H. This
basis is the so-called ``adiabatic basis'', in the sense that the states
``follow'' the field, and is obtained by means of the unitary transformation 
\begin{equation}
U_{D\rightarrow A}=\left( 
\begin{array}{ll}
\cos \chi  & \sin \chi  \\ 
-\sin \chi  & \cos \chi 
\end{array}
\right) ,  \label{rotation}
\end{equation}
with $\chi =-1/2\arctan (2x_{10}E(t)/\omega _{10}).$ This gives 
\begin{equation}
H^{A}=U_{D\rightarrow A}HU_{D\rightarrow A}^{T}=\left( 
\begin{array}{ll}
\varepsilon^A _{-} & 0 \\ 
0 & \varepsilon^A _{+}
\end{array}
\right) ,
\label{Hadiab}
\end{equation}
where the field-dressed energies are given by
\begin{equation}
\varepsilon^A _{\pm }=\pm \frac{1}{2}\sqrt{\omega _{10}^{2}+(2x_{10}E(t))^{2}}.
\label{edress}
\end{equation}
Applying $U_{D\rightarrow A}$ to the diabatic basis states, one obtains the
field-dressed, ``adiabatic'' states 
\begin{equation}
\left| \phi _{0}^{A}(t)\right\rangle =\cos \chi \left| \phi
_{0}^{D}\right\rangle +\sin \chi \left| \phi _{1}^{D}\right\rangle 
\end{equation}
and 
\begin{equation}
\left| \phi _{1}^{A}(t)\right\rangle =-\sin \chi \left| \phi
_{0}^{D}\right\rangle +\cos \chi \left| \phi _{1}^{D}\right\rangle ,
\end{equation}
whose energies are, respectively, $\varepsilon^A _{-}$ and 
$\varepsilon^A _{+}$ \cite{SuomG92}.
In order to compute the harmonic spectra, one needs the Fourier transform of
the time-dependent dipole. This quantity is given, in its length and
acceleration form, by 
\begin{equation}
x=x_{10}\left[ g(t)\cos 2\chi +h(t)\sin 2\chi \right],   \label{d1}
\end{equation}
and
\begin{equation}
\ddot{x}=-\omega _{10}^{2}x+2\omega _{10}x_{10}^{2}E(t)\left[ h(t)\cos
2\chi -g(t)\sin 2\chi \right],   \label{d2}
\end{equation}
respectively,
with $g(t)=C_{0}^{*A}(t)C_{1}^{A}(t)+C_{1}^{*A}(t)C_{0}^{A}(t)$ and $%
h(t)=|C_{0}^{A}(t)|^{2}-|C_{1}^{A}(t)|^{2},$ where $C_{n}^{A}(t)=\left\langle
\phi _{n}^{A}(t)\right. \left| \psi (t)\right\rangle $ denotes the
projection of the wave function $\left| \psi (t)\right\rangle$ onto an
adiabatic state. The equations above are the superposition of two distinct
terms, namely the crossed terms and the population difference between the
adiabatic states. Since the population difference $h(t)$ roughly ``follows"
the field, it contributes mainly to the generation of low harmonics, 
whereas $g(t)$ is expected to be responsible for the high harmonics. This has
been confirmed by numerical studies (not shown).

An interesting feature is that, in the extreme limit $E_{0}\rightarrow
\infty $, the transformation (\ref{rotation}) formally corresponds to (\ref
{rotation2}) and the dipole length (\ref{d1}) becomes proportional to
 the population
difference between the adiabatic states. However, one should keep in mind
that, only in this limit, the states obtained using (\ref{rotation2}) on the
field-free states and the adiabatic states are formally equivalent. In
general, this is not the case.

In the subsequent sections, we work mainly in the adiabatic basis, and refer to
crossings of the exchanged-basis energies. For the adiabatic energies, there
are avoided crossings. The results
discussed in this paper have been obtained from the numerical solution 
of Eq. (\ref{twolev})
in the adiabatic basis,  by means of
a fourth-order Runge-Kutta method. Unless stated otherwise, the driving field
is turned on instantaneously.

\subsection{Windowed Fourier transform}

For both open and closed systems, high-harmonic generation
is always related to abrupt population transfers. Depending on the group of
harmonics, they occur at particular times, which give the main contributions
to high-harmonic generation within a field cycle. For an atom in a strong
laser field, for instance, these times are well-known and correspond to the
return times $t_{1}$ of an electron which left an atom at a previous time $%
t_{0}$. For a closed system, the times $t_{0}$ correspond to the
level-crossing times and the times $t_{1}$ are still an open question to
some extent. A very useful method to extract these latter times from the
time-dependent dipole is performing a Fourier transform with a temporally
restricted window function. For an arbitrary function $f(t^{\prime })$, this
transform is

\begin{equation}
{\cal F}(t,\Omega ,\sigma )=\int\limits_{-\infty }^{+\infty }dt^{\prime
}f(t^{\prime })W(t,t^{\prime },\Omega ,\sigma )\quad ,
\end{equation}
where $t,\Omega $ and $\sigma $ denote the time and harmonic frequency at
which the window function is centered, and its temporal width,
respectively. We consider a Gabor transform, for which the window function is
given by 
\begin{equation}
W(t,t^{\prime },\Omega ,\sigma )=\exp [-(t-t^{\prime })\sp 2/\sigma
^{2}]\;\exp [{\rm i}\Omega t^{\prime }]\ .
\label{window}
\end{equation}
The usual Fourier transform ${\cal F}(\Omega ),$ which yields no temporal
information, is recovered for $\sigma \rightarrow \infty $. The temporal
width $\sigma $ corresponds to a frequency bandwidth $\sigma _{\Omega
}=2/\sigma $. For a temporal width smaller than the period $T=2\pi /\omega $
of the driving field, the peaks in the time-resolved spectra
$|{\cal F}(t,\Omega ,\sigma )|^2$ yield the
recombination times $t_{1}.$ This method has been extensively used in the
literature, in the three-step model framework \cite{tprof}.

\section{General picture}
\label{res1}

We shall now investigate the connection between HHG and the periodic level
crossings in detail and draw a general physical picture of the 
mechanisms involved.
The simplest physical situation for which one can do this is a monochromatic field 
\begin{equation}
E(t)=E_{0}\sin(\omega t),
\end{equation}
where $E_{0}$ and $\omega$ denote the field strength and frequency,
respectively. In this case, the time $t_c$ is given by 
the condition
\begin{equation}
\omega t_c=\frac{\omega_{10}}{2x_{10}E_0}.
\end{equation}
If the field amplitude $E_0$ is large enough, then 
$\omega t_c \ll 1$, and the avoided crossings of the adiabatic 
states are well-separated. 
Thus, the crossing times $t_0$ are well-defined and there is efficient 
population transfers at $t_0$. Hence,
 one expects the corresponding
spectrum to exhibit a wide plateau and a sharp cutoff.

The avoided crossings occur at the times $t_{0}=n\pi /\omega $
for which the field is vanishing. Thus, one expects the population transfers
between the states $\left| \phi _{n}^{A}(t)\right\rangle $ to occur at these
times. This is partially confirmed by Fig. 1, where the populations of the
adiabatic states are plotted as functions of time. In fact, the pronounced
peaks at the times $t_{0}$ clearly show that most population transfer takes
place at these times. There are however several smaller peaks, which are
symmetric with respect to the times $t_{1M}=(2n+1)\pi /2\omega $ for which
the field is maximal. These peaks show that population transfer also occurs
at other times, and can be seen in detail in Fig. 1(b). 
\begin{figure}[tbp]
\begin{center}
\epsfig{file=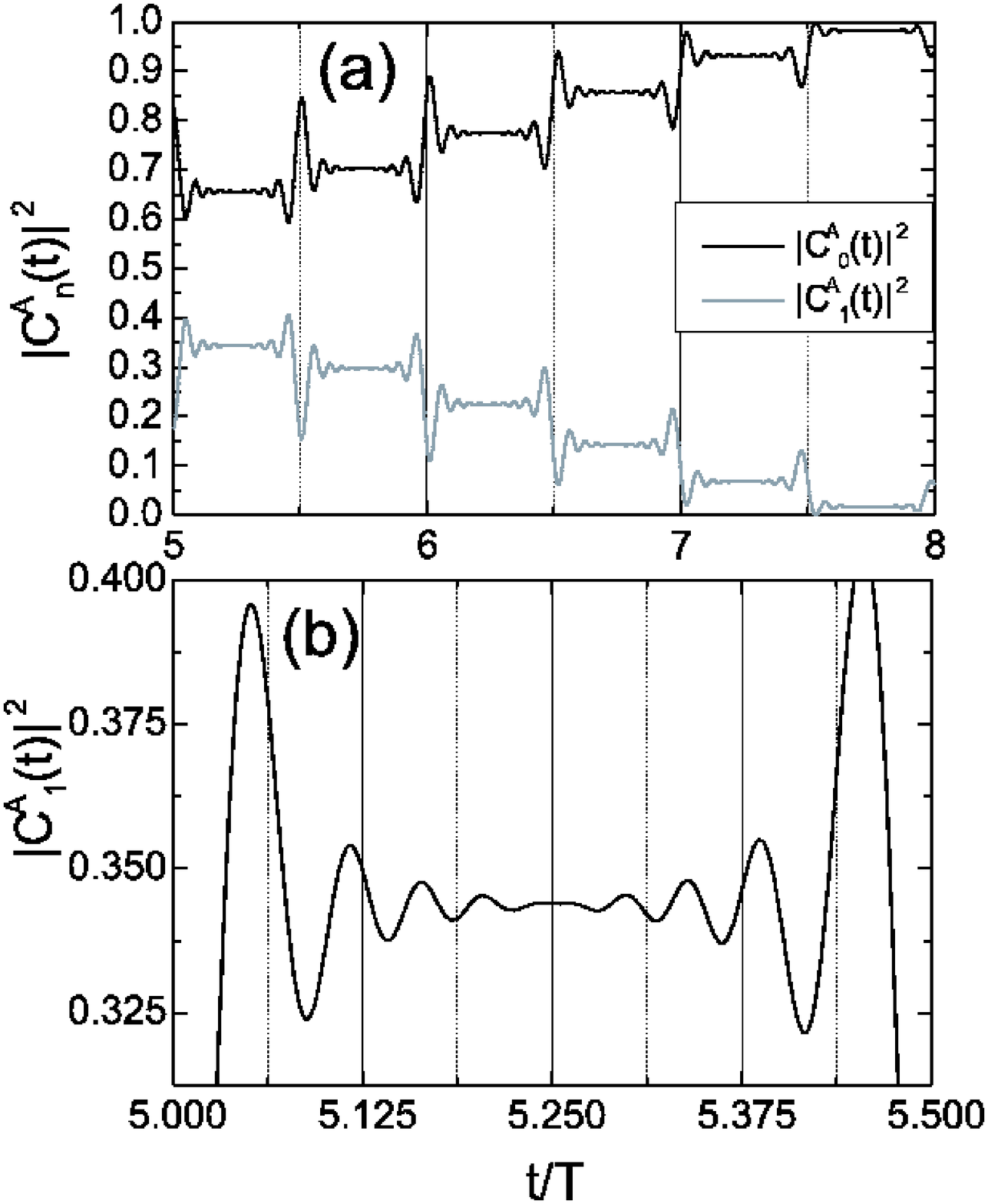,width=7.0cm,angle=0}
\end{center}
\caption{Populations $|C_{n}^{A}(t)|^{2}$ of the adiabatic states as
functions of time, for transition frequency $\omega _{10}=0.409$ a.u.,
external field parameters $\omega =0.05{\rm a.u.}$, $E_{0}=0.6%
$ a.u., and dipole-matrix element $x_{10}=1.066$ a.u.. Part (a) shows this 
feature for several cycles of the driving field, whereas part (b) depicts 
the population of the excited adiabatic state only within half a cycle. The
times are given in units of the field cycle $T=2\pi/\omega$. The driving field
is turned on linearly within two periods.}
\end{figure}

The role of these population transfers in HHG can be understood using the
Gabor transform of the dipole acceleration. The peaks
in the Gabor spectra give the main contributions for high-harmonic
generation within a field cycle. For the cutoff harmonic, there is a single
peak at $t_{1M}$ which splits into two, for the plateau harmonics. This peak
gets further apart as the harmonic frequency decreases, varying from $t_{1M}$
to the times at the immediate vicinity of the avoided crossings. These 
results are displayed in Fig. 2.

\begin{figure}[tbp]
\begin{center}
\epsfig{file=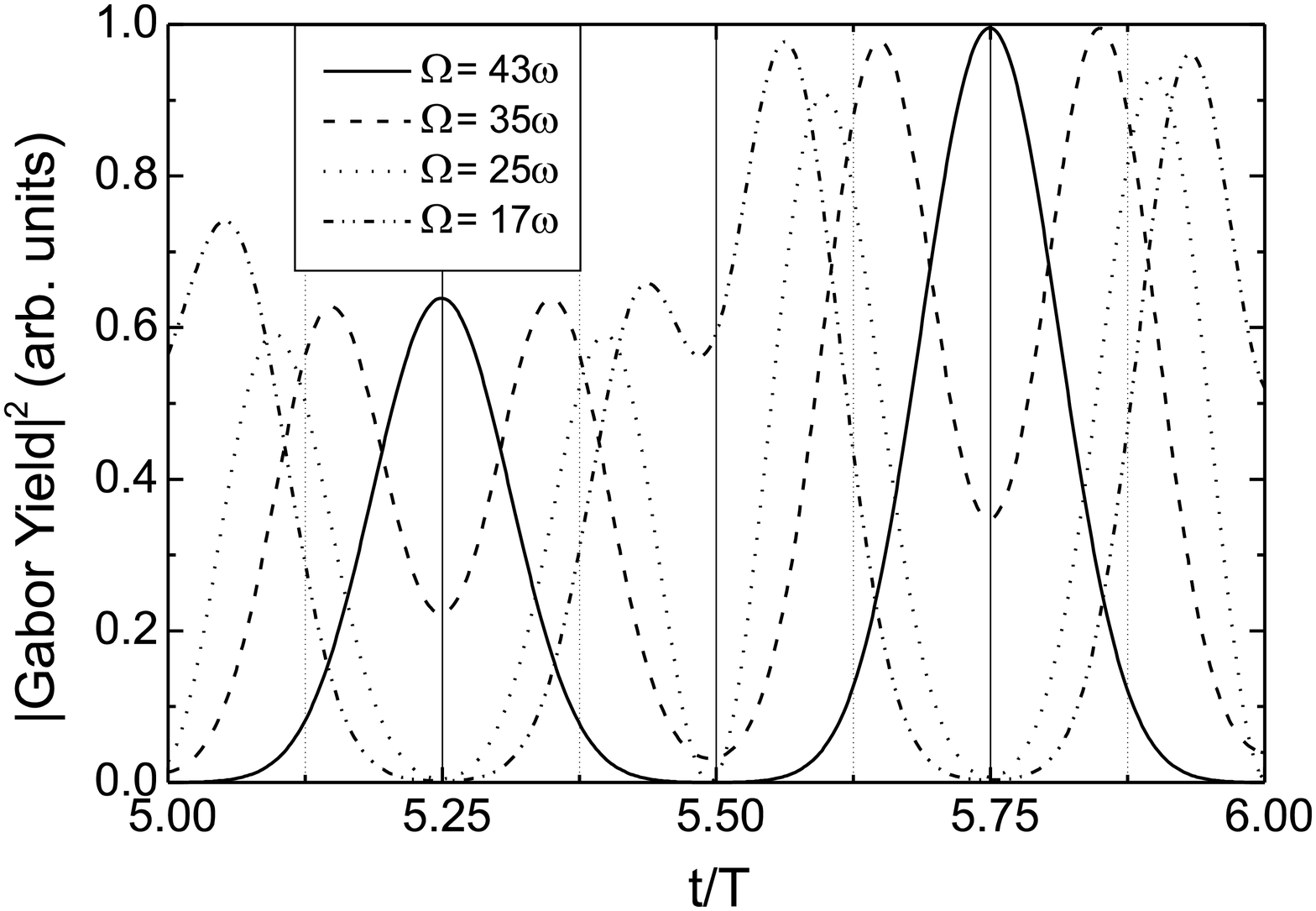,width=7.0cm,angle=0}
\end{center}
\caption{Gabor spectra of the dipole 
acceleration (Eq. (11)) as functions of time, for field
strength $E_0=1$ a.u., field frequency $\omega=0.05$ a.u., 
transition frequency $\omega_{10}=0.409$ a.u. and dipole matrix 
element $x_{10}=1.066$ a.u.. 
The cutoff harmonic lies at
$\Omega_M=43\omega$. The time width of the window 
function was chosen $\sigma =0.1T$. Its center was chosen at the cutoff
harmonics, as well as at harmonic energies which roughly correspond to 
$\Omega=0.8\Omega_M$, $\Omega=0.6\Omega_M$, and
$\Omega=0.4\Omega_M$. All time-resolved spectra have been normalized. The
times are given in units of the field cycle $T=2\pi/\omega$. The driving field
is turned on linearly within two periods.}
\end{figure}

The physical interpretation of these features is rather simple. At the times
the level crossings occur, i.e., at $t_{0}=nT/2$, there is a population
transfer from the adiabatic state $\left| \phi _{0}^{A}(t)\right\rangle $ to 
$\left| \phi _{1}^{A}(t)\right\rangle .$ The system remains in $\left| \phi
_{1}^{A}(t)\right\rangle $ until a further time $t_{1}$, decaying back to $%
\left| \phi _{0}^{A}(t)\right\rangle $ and emitting a harmonic of frequency $%
\Omega =N\omega=\varepsilon^A _{+}-\varepsilon^A _{-}.$ The explicit expression relating
the time $t_{1}$ to the harmonic frequency would then be 
\begin{equation}
\omega t_{1}=\arcsin \left[ \pm \sqrt{(N\gamma_1) ^{2}-(\gamma_2)^{2}%
}\right] ,  \label{tret}
\end{equation}
with $\gamma_1=\omega/(2x_{10}E_0)$ and
$\gamma_2=\omega_{10}/(2x_{10}E_0)$. The physical significance of $\gamma_1$
and $\gamma_2$ will be discussed later in this paper (Sec. \ref{res3}). 
In order to obtain a harmonic at the maximum possible frequency 
$\Omega_M$ (i.e., the cutoff harmonic), the population transfer 
between the time-dependent states
must occur at the times for which the energy difference $\varepsilon^A
_{+}-\varepsilon^A _{-}$ is maximal, i.e., at $t_{1M}=(2n+1)\pi /2\omega $.
 As the harmonic energy decreases, there are two possible times for this
population transfer to occur, a shorter and a longer one. The interference
between these two possible quantum paths originates the two-level atom
plateau. This process repeats itself every half cycle of the driving field.
This picture is supported by the fact that all peaks in the time-resolved 
spectra satisfy Equation (\ref{tret}) and thus can be traced back 
to population 
transfers
between the adiabatic states. The times given by (\ref{tret}) for the
parameters of Fig. 2, together with
the corresponding harmonic energies, are written in Table I. 

An analogous picture is observed within the three-step model framework. The
cutoff harmonic can only be generated by an electron which returns to its
parent ion with maximal kinetic energy. This maximal energy corresponds to a
particular return time, which appears as a single peak in the Gabor yield. 
Within the plateau, there are two possible sets of
electron trajectories corresponding to the same harmonic energy, such that
this single peak splits into two \cite{tprof}. In our case, the ``first
step'' would be the population transfer from $\left| \phi
_{0}^{A}(t)\right\rangle $ to $\left| \phi _{1}^{A}(t)\right\rangle $ at $%
t_{0},$ the ``second step'' would be the system following 
$\left| \phi _{1}^{A}(t)\right\rangle $ adiabatically in a time interval
$\tau=t_1-t_0$ and the ``third step'' the population
transfer from $\left| \phi _{1}^{A}(t)\right\rangle $ to $\left| \phi
_{0}^{A}(t)\right\rangle $ at $t_{1}$, with subsequent harmonic generation.
The corresponding physical picture is illustrated in Fig. 3. 
\begin{figure}[tbp]
\begin{center}
\epsfig{file=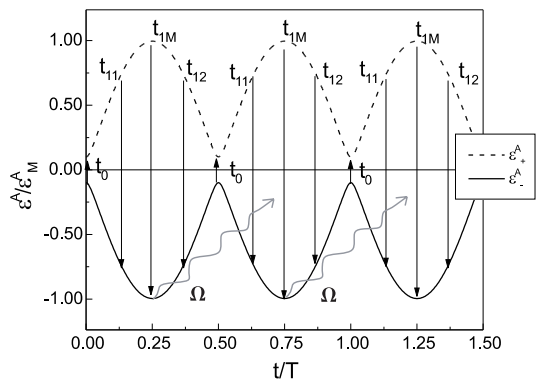,width=7.0cm,angle=0}
\end{center}
\caption{Schematic representation of high-harmonic generation in a two-level
atom. The population transfers at the level crossings occur at the times $%
t_{0}$ and the main contributions to HHG occur at the times $t_{1}.$ The
times $t_{1M}$, $t_{11}$ and $t_{12}$ correspond to the generation of the
cutoff and plateau harmonics, respectively. The main physical processes are
indicated by arrows in the figure, and the corresponding energies can be
read in the vertical axis. The adiabatic energies are given in units of the
maximal energy $\varepsilon^A_M$ and the time in units of the field cycle. The
field parameters are chosen in such a way that the ratio between the cutoff
energy $\Omega_M=2\varepsilon^A_M$ and the transition frequency is $\Omega_M/\omega_{10}=10.$}
\end{figure}

Another interesting feature is that the population transfers between the
adiabatic states are not strictly periodic within $\pi /\omega $. Indeed, 
superposed to them, there are oscillations which occur within much larger
time scales, their periods comprising several cycles of the driving field 
\cite{GaVit97}. These oscillations are also present in the dipole length and
acceleration as a gobal enveloping function, whose amplitude, form and
periodicity depend on the field strength $E_{0}$, the field frequency $%
\omega $ and on the dipole matrix element $x_{10}$ in a non-trivial way.
These structures seem not to influence the harmonics gobally, but mainly the
substructure of the spectra and the hyper-Raman lines \cite{footnote0}. 

In Fig. 4, we show these enveloping
functions for the populations of the adiabatic states (Fig. 4(a)), the
dipole acceleration (Fig. 4(b)), and the Gabor spectra of the plateau and
cutoff harmonics (Fig. 4(c)). One should note that this enveloping function
is the same for the Gabor transforms of all groups of harmonics displayed.
Furthermore, it does not affect the splitting of the peaks, such that the
population transfer times are always given by Eq. (\ref{tret}). 
\begin{figure}[tbp]
\begin{center}
\epsfig{file=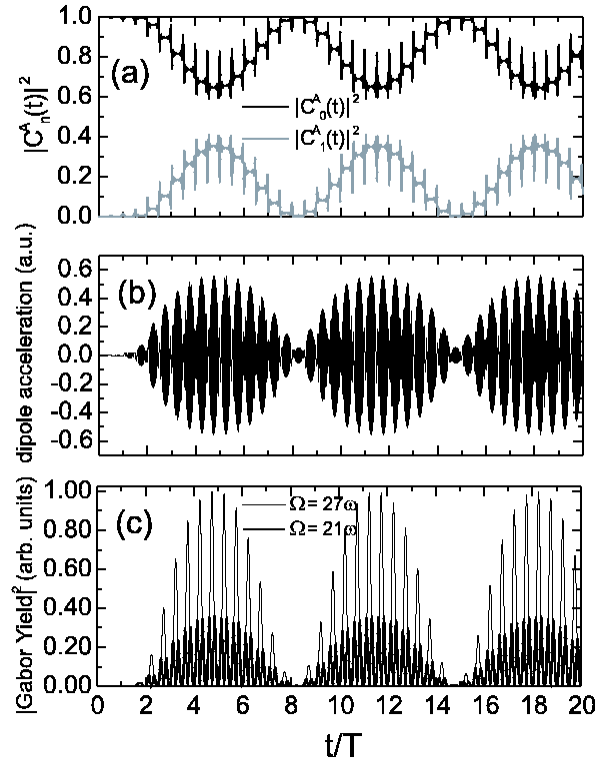,width=8.0cm,angle=0}
\end{center}
\caption{Global structures as functions of time, for: (a) the populations $%
|C_{n}^{A}(t)|^{2}$ of the adiabatic states; (b) the dipole acceleration $%
\ddot{x}(t)$; (c) the Gabor spectra of the cutoff and plateau harmonics.
 The time width of the window function is $\sigma=0.1T$. The field strength, 
 the field frequency, the transition frequency and the
dipole matrix element were chosen as $E_0=0.6$ a.u., $\omega=0.05$ a.u., 
 $\omega_{10}=0.409$ a.u. and $x_{10}=1.066$ a.u., respectively. 
These parameters give $\gamma_1=0.0391$, $\gamma_2=0.3197$ and a cutoff
frequency at $\Omega_{\max}=27\omega$. All Gabor spectra have been normalized
to 
the maximum value obtained with the window function centered at the
cutoff. The field is turned on linearly within two periods. 
The time is given in units of the field cycle.}
\end{figure}

\section{Control}

\label{res2}In this section, we consider a bichromatic driving field 
\begin{equation}
E(t)=E_{01}\sin (\omega t)+E_{02}\sin (n\omega t+\theta ),  \label{bichr}
\end{equation}
with two main purposes. First, we wish to confirm the physical picture in
which the main contributions to a particular set of harmonics, within a
field cycle, occur at the times $t_{1}$ such that the corresponding
harmonic frequency is the difference $\varepsilon^A _{+}-\varepsilon^A _{-}$
between the energies of the adiabatic states. Second, we are interested in
understanding how an additional field can be used to distort the avoided
crossings between the adiabatic states in such a way that the harmonic
emission can be controlled. In the bichromatic case, depending on the field
parameters, the spectra may have several cutoffs, which are given by the
maxima of $\varepsilon^A _{+}-\varepsilon^A _{-}.$ Consequently, the main
contributions to the generation of the cutoff harmonics take place at the
times $t_{1M}$ for which these maxima occur.

In order to obtain the level-crossing times $t_{0}$, as well as the times $%
t_{1M}$, one needs the extrema $\pm \varepsilon^A_M$ of the field-dressed 
energies $\varepsilon^A_{\pm }.$ For the bichromatic field (\ref{bichr}) 
they are given by 
\begin{equation}
\cos (\omega t)+n\zeta \cos (n\omega t+\theta )=0  \label{maxima}
\end{equation}
and 
\begin{equation}
\sin (\omega t)+\zeta \sin (n\omega t+\theta )=0,  \label{cross}
\end{equation}
where $\zeta =E_{02}/E_{01}$ denotes the field-strength ratio. Equation (\ref
{maxima}) gives the extrema which coincide with those of the field, and
therefore $t_{1M},$ whereas Equation (\ref{cross}) gives those which
correspond to the avoided crossings, and therefore $t_{0}$. Depending on the
frequency ratio $n$, the field-strength ratio $\zeta $ and the relative
phase $\theta ,$ these times, as well as the corresponding extrema, can be
very different. In this paper, we will provide concrete examples for a $%
\omega -2\omega $ field, i.e., with $n=2$, relative phases $\theta _{1}=0$
and $\theta _{2}=\pi /2,$ and arbitrary $\zeta$. For these specific
parameters, (\ref{maxima}) and (\ref{cross}) have a simple form, with
analytical solutions.

\subsection{$\theta =0$}

In this case, Eq. (\ref{maxima}) reduces to 
\begin{equation}
\cos ^{2}(\omega t)+\frac{1}{4\zeta }\cos (\omega t)-\frac{1}{2}=0,
\end{equation}

which yields two sets of times, namely 
\begin{equation}
t_{1M}=\frac{1}{\omega }\arccos \left( -\frac{1}{8\zeta }\pm \frac{1}{2}%
\sqrt{\frac{1}{16\zeta ^{2}}+2}\right) .  \label{tmaxi}
\end{equation}
The solutions corresponding to the positive root exist for all
field-strength ratios, whereas the remaining solutions are only present for $%
\zeta >0.5$. Further in this section, it will be shown that the first set
gives the absolute maxima of $\varepsilon^A _{\pm }$, which correspond to the
cutoff in the harmonic spectra, whereas the second set yields local maxima
at much lower energies.

The expression giving the avoided crossings, on its turn, can be written as 
\begin{equation}
\sin (\omega t)\left[ 1+2\zeta \cos (\omega t)\right] =0.
\label{crossbic1}
\end{equation}
This equation yields the crossing times $t_{0}=n\pi /\omega $, and 
$t_{0}^{\prime }=1/\omega\hspace*{0.1cm} \arccos [-1/(2\zeta )].$ The crossing times $%
t_{0}$ do not depend on the field-strength ratio and are the same as in the
monochromatic case, whereas the crossing times $t_{0}^{\prime }$ clearly do.
Furthermore, these latter times are only present for $\zeta >0.5.$%
\begin{figure}[tbp]
\begin{center}
\epsfig{file=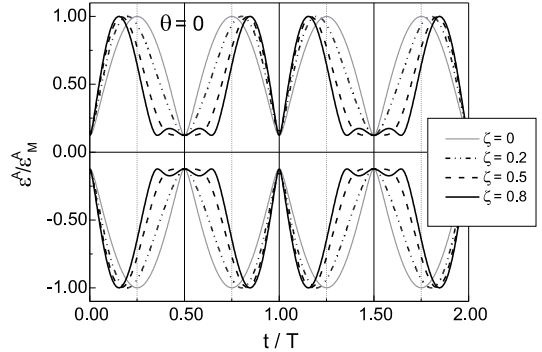,width=7.0cm,angle=0}
\end{center}
\caption{Energies of the adiabatic states for a bichromatic field $%
E(t)=E_{01}\sin (\omega t)+E_{02}\sin (2\omega t+\theta ),$ for $\theta =0$
and several field-strength ratios $\zeta =E_{02}/E_{01}.$ The time $t$ is
given in units of the field cycle $T=2\pi/\omega$ and the field-dressed 
energies in units of
the maximal energy $\varepsilon^A_M$. The field parameters were chosen such
that $\Omega_M/\omega _{10}=8.$ }
\label{bic_1}
\end{figure}

Fig. \ref{bic_1} gives concrete examples of how the adiabatic energies $%
\varepsilon^A _{\pm }$ depend on time, for different field-strength ratios.
In contrast to the monochromatic
case,  $\varepsilon^A_\pm$ is not periodic within half a
cycle of the driving field. This is not surprising, since the periodicity of
the field-dressed energies is effectively determined by $E^2(t)$ 
(c.f. Eq. (\ref{edress})). 
For a monochromatic field,
$E^2(t)=E^2(t+\pi/\omega)$ always holds, whereas in the bichromatic case this
is only true for odd frequency ratios $n$. This is clearly {\it not} the case
addressed in this paper. For the phase $\phi=0$, one observes that
$\varepsilon^A_\pm(t)=\varepsilon^A_\pm(2\pi/\omega-t)$, if both times are
taken symmetrically with respect to $t_0=n\pi/\omega$. This property already
reflects itself in the expressions for $t_0$, $t_{1M}$ and $t'_0$ derived in this section.

Furthermore, one clearly sees that, as predicted in Eq. (\ref{crossbic1}), 
for $\zeta <0.5,$ 
the second driving
wave only distorts the avoided crossings, making them broader at $%
t_{0}=(2n+1)\pi /\omega $ and sharper at $t_{0}=2n\pi /\omega .$ For $\zeta
=0.5$, the broad crossing starts to split, originating the crossings given
at the times $t_{0}^{\prime }.$ This splitting also leads to the second set
of maxima predicted by Eq. (\ref{tmaxi}), which corresponds to a set of
harmonics of relatively low frequencies. 

One must now understand which consequences this effect has on the physical
quantities involved. With that purpose, we choose the strengths of both driving waves such
that $\varepsilon^A_{M}$, and therefore the cutoff energy,
 remains unchanged and is
equal to the monochromatic cutoff
energy, for variable field-strength ratio $\zeta$. This gives 
\begin{equation}
E_{01}=\frac{E_0}{\sqrt{1-\beta^2}(1+2\beta\zeta)},
\label{strength}
\end{equation}
with $\beta=\cos(t_{1M})$.
\begin{figure}[tbp]
\begin{center}
\epsfig{file=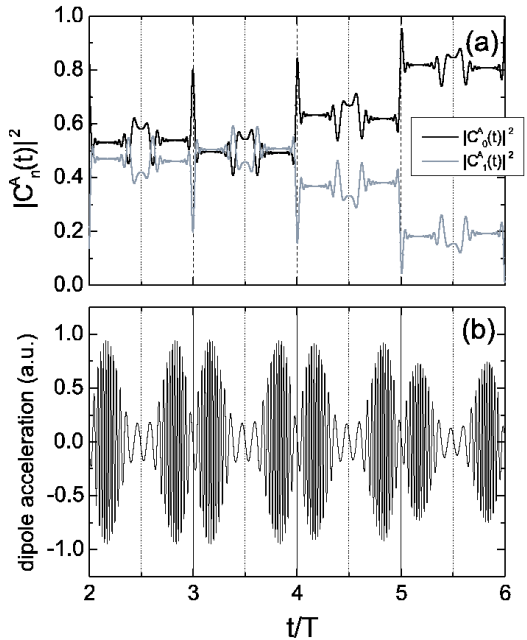,width=7.0cm,angle=0}
\end{center}
\caption{Populations $|C^A_n(t)|^2$ of the adiabatic states (Part (a)) and 
dipole acceleration (Part (b)) as 
functions of time, for a bichromatic field $
E(t)=E_{01}\sin (\omega t)+E_{02}\sin (2\omega t+\theta ),$ with $\theta =0$,
$\omega=0.05$ a.u., $\omega_{10}=0.409$ a.u., $x_{10}=1.066$
a.u. and field-strength ratio $\zeta =E_{02}/E_{01}=0.5.$ The field amplitudes
were chosen according to Eq. (\protect{\ref{strength}}), with $E_0=1$ a.u.. The time $t$ is
given in units of the field cycle.}
\label{bic_2}
\end{figure}

The population transfers between the adiabatic states, as functions of time,
also exhibit very similar asymmetries 
to the ones observed in the field-dressed energies. The population transfers
at the broad crossings, for instance, take place at longer time intervals
than those at the sharp crossings, making the oscillations in $|C^A_n(t)|^2$
asymmetric with respect to the times $t_{1M}$. This asymmetry increases with
increasing $\zeta$. An example is provided in Fig. \ref{bic_2}(a).
 A similar feature occurs for the dipole acceleration. 
This highly oscillating function exhibits nodes at the
level-crossing times. In the monochromatic case, these nodes extend over
identical temporal
 regions every half-cycle of the driving field. For bichromatic fields,
however,
with the distortion of the crossings by the second
driving wave, this picture changes. There exist
narrower and broader nodal regions, corresponding to the narrower and broader
crossings, respectively. Thus, the oscillations of the dipole acceleration
get ``squeezed" between the broader nodes. This feature can be seen in Fig.
 \ref{bic_2}(b).

\begin{figure}[tbp]
\begin{center}
\epsfig{file=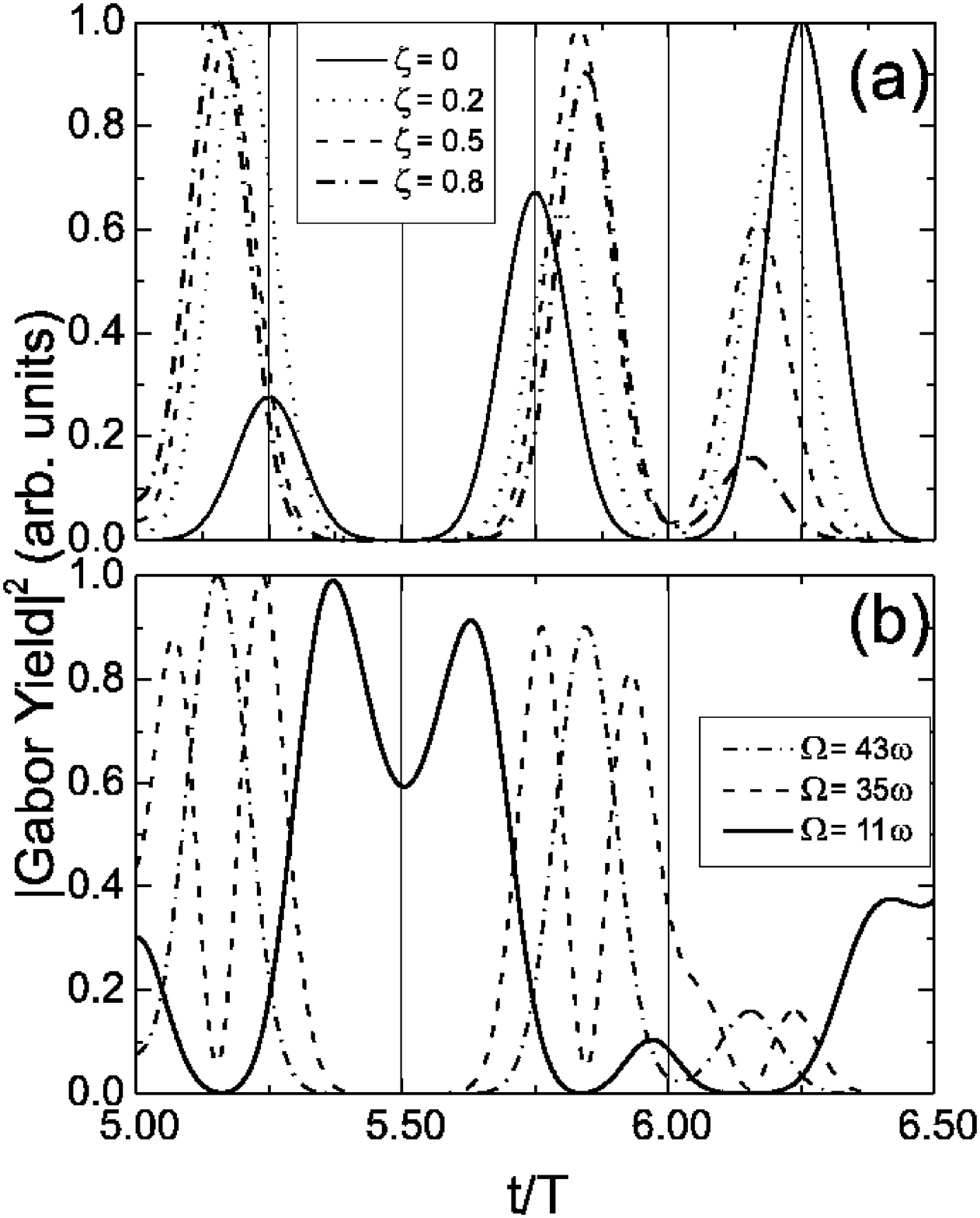,width=7.0cm,angle=0}
\end{center}
\caption{Gabor spectra of the dipole acceleration as functions of time, for
a bichromatic field 
$E(t)=E_{01}\sin (\omega t)$ $+E_{02}\sin (2\omega t+\theta ),$ with $\theta =0$,
$\omega=0.05$ a.u., $\omega_{10}=0.409$ a.u., $x_{10}=1.066$
a.u. and several field-strength ratios $\zeta =E_{02}/E_{01}.$ The maximal
field strength is kept fixed according to Eq. ({\protect\ref{strength}}), with
$E_{0}=1$ a.u. The cutoff energy
lies at $\Omega_{M}=2\varepsilon^A_M=43\omega$. The temporal width
of the window function is $%
\sigma =0.1T$. In Part (a), the window function is centered at the cutoff
harmonics, and the field-strength ratio is $0\leq \zeta \leq 0.8.$ In Part
(b), the center of the window function is taken for different
frequencies, and $\zeta =0.8$. All curves in the figure have been normalized
to their maximum values. }
\label{gabbic1}
\end{figure}

The Gabor transform of the dipole acceleration, taken at the cutoff and in
the plateau, confirms this picture. In
Fig. \ref{gabbic1}(a), there is a clear displacement of the peaks in the
time-resolved spectra for the cutoff harmonics, with respect to the monochromatic case,
and these peaks occur at the times predicted by Eq. (\ref{tmaxi}). Similarly
to the monochromatic case, these peaks split into two in the plateau region,
being, however, slightly asymmetric (Fig. \ref{gabbic1}(b)). This asymmetry is related
to the above-mentioned difference in the shapes of the crossings. 
Furthermore, for larger
field-strength ratio, the additional times can also be seen, for a group of
harmonics at the low-energy end of the plateau. The times $%
t_{0}$ and $t_{1M}$, together with the respective cutoff energies, are given in
Table II for the specific parameters considered in this figure.

\subsection{$\theta =\pi /2$}

For this relative phase, Equation (\ref{maxima}) has the form 
\begin{equation}
\cos (\omega t)\left[ 1-2\zeta \sin (\omega t)\right] =0.  \label{maxi2}
\end{equation}
This equation has two types of solutions: $t_{1M}=(n+1/2)\pi /\omega $,
which do not depend on the field-strength ratio and yield the same maxima as
in the monochromatic case, and $t_{1M}^{\prime }=1/\omega \arcsin [1/(4\zeta
)],$ which clearly depend on $\zeta $ and exist only for $\zeta \geq 0.25$.
This already hints at a completely different situation as in the previous
section, which will now be discussed in detail. This also holds for the
times at which the avoided crossings occur. They must now satisfy 
\begin{equation}
\sin ^{2}(\omega t)-\frac{1}{2\zeta }\sin (\omega t)-\frac{1}{2}=0
\end{equation}
such that 
\begin{equation}
t_{0}=\frac{1}{\omega }\arcsin \left( \frac{1}{4\zeta }\pm \frac{1}{2}\sqrt{%
\frac{1}{4\zeta ^{2}}+2}\right) ,  \label{cross2}
\end{equation}
all of them depending on $\zeta .$ This means that, in contrast
to the case $\theta =0,$ one may shift all level-crossing times by changing the
relative intensities of the driving waves. The set of crossings given by the
positive root in (\ref{cross2}) exists only for $\zeta \geq 1,$ whereas the
remaining crossings occur for all $\zeta .$
 
\begin{figure}[tbp]
\begin{center}
\epsfig{file=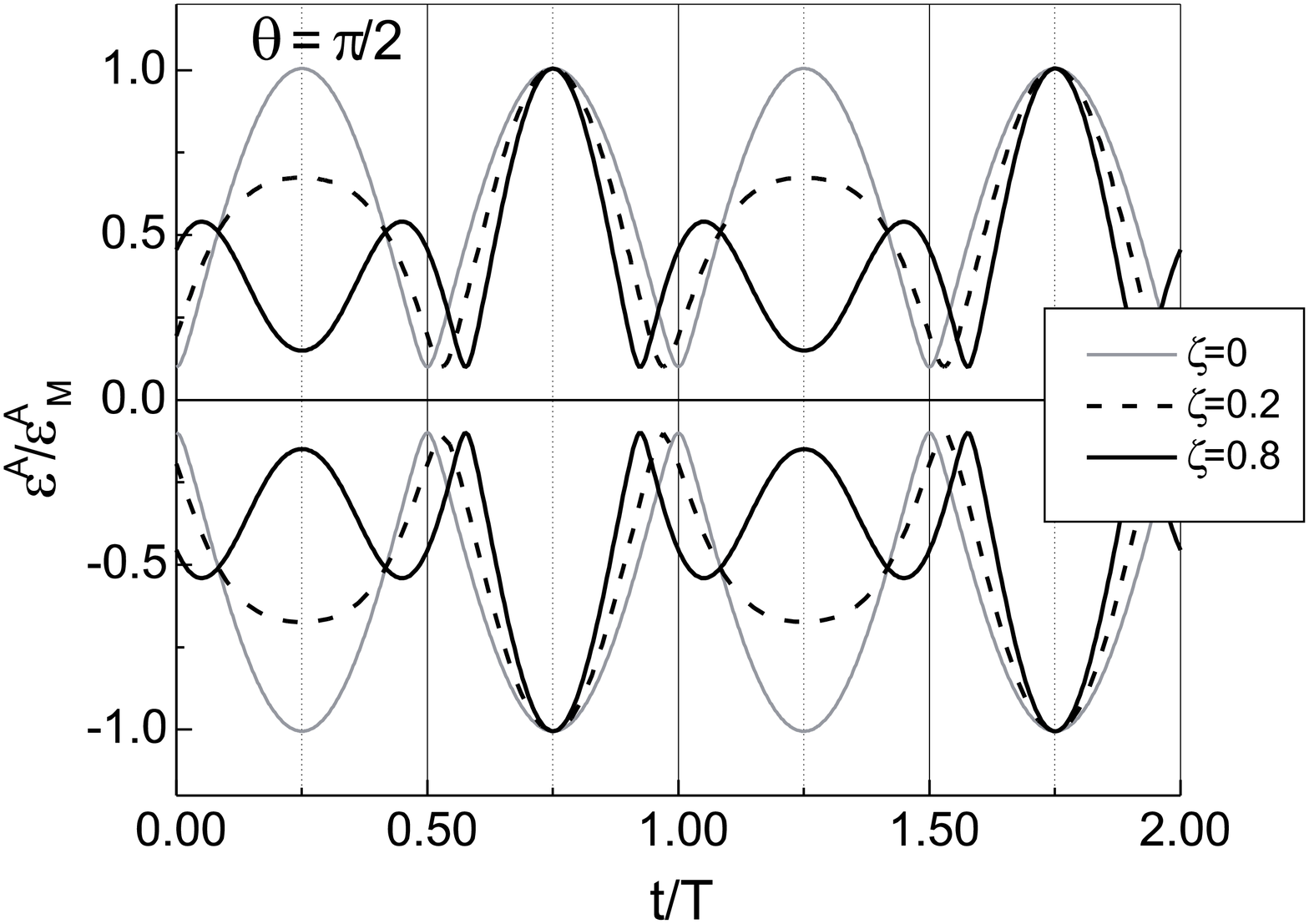,width=7.0cm,angle=0}
\end{center}
\caption{Energies of the adiabatic states for a bichromatic field $%
E(t)=E_{01}\sin (\omega t)+E_{02}\sin (2\omega t+\theta ),$ for $\theta =\pi
/2$ and several field-strength ratios $\zeta =E_{02}/E_{01}.$ The time $t$
is given in units of the field cycle and the field-dressed energies in units
of the maximal energy $\varepsilon^A_{M_2}$. The field parameters were chosen such that $\Omega_{M_2}/\omega _{10}=8.$ The times $t_{1M_i}$ are 
indicated in the figure by the dotted and solid grid lines, respectively.}
\label{bic_3}
\end{figure}

In Fig. \ref{bic_3}, we depict the adiabatic states as functions of time,
for several values of $\zeta ,$ similarly to what was done for $\theta =0.$
This figure illustrates how the relative phase can radically alter the whole
physical picture. For $\theta =\pi /2$, already a relatively weak high-frequency
wave considerably distorts the avoided level crossings, as well as the
maxima of the field-dressed energies. An interesting feature is that the
avoided crossings now move with the field-strength ratio. Furthermore, the
maximal energies are no longer equal, but, within a field cycle, there are two
comparable and different cutoff energies. This can be directly seen by
computing the extrema of the energies $\varepsilon^A _{\pm }$, which occur for 
$t_{1M}.$ 

For field-strength ratio $\zeta<0.25,$ they give the energies
\begin{equation}
\varepsilon^A _{M_1}=\frac{1}{2}\sqrt{\omega
_{10}^{2}+4x_{10}^{2}(E_{01}-E_{02})^{2}}
\end{equation}
 and 
\begin{equation}
\varepsilon^A _{M_2}=\frac{1}{2}\sqrt{\omega
_{10}^{2}+4x_{10}^{2}(E_{01}+E_{02})^{2}},
\end{equation}
which correspond to the times
$t_{1M_1}=0.25T \hspace*{0.1cm}\mathrm{mod} \hspace*{0.1cm}T$, and to
 $t_{1M_2}=0.75T\hspace*{0.1cm} \mathrm{mod}\hspace*{0.1cm} T$, respectively.
These times define symmetry axes for the time-dependence of 
the adiabatic energies.

 For $\zeta \geq 0.25$, a further splitting of the set of maxima at 
$t_{1M_1}$ occurs, as predicted
in Eq. (\ref{maxi2}). There exist now two sets of maxima, at the
 times $t'_{1M}$, whose energies are equal and given by
\begin{equation}
\varepsilon^A _{M_1}=\frac{1}{2}\sqrt{\omega
_{10}^{2}+4x_{10}^{2}E_{01}^{2}\frac{(1+8\zeta^2)^2}{64\zeta^2}}.
\end{equation}
These maxima are symmetric with respect to $t_{1M_1}$. For these times, the
adiabatic energies now exhibit a minimum. This causes, for large $\zeta$, 
additional avoided crossings (c.f. Fig. 8 for $\zeta=0.8$). The population
transfers at these times are however small, and play only a secondary role in
the problem addressed in this paper. For the sake of simplicity, even after the
second splitting, we shall refer to the lower-energy set of maxima as 
$\varepsilon^A_{M_1}$.
 The other set of maxima does not split, and the corresponding times $t_{1M_2}
$ remain constant for all $\zeta$.
One should note that the adiabatic energies, in the $\theta=\pi/2$ 
case, satisfy $\varepsilon^A_\pm(t)=\varepsilon^A_\pm(T/2-t)$, if both times
are chosen symmetrically with respect to $t_{1M_1}$ or $t_{1M_2}$. This also holds
for the population-transfer times derived in this section.

In order to investigate how the distortions in the adiabatic-state energies
influence the physical quantities of interest, we shall keep the cutoff energy
$\Omega_{M_2}=2\varepsilon^A_{M_2}$ fixed, and equal to the cutoff energy of
the monochromatic case. Thus, the field strengths $E_{01}$ and 
$E_0$ are related by
\begin{equation}
E_{01}=\frac{E_0}{1+\zeta}
\label{strength2}
\end{equation}

\begin{figure}[tbp]
\begin{center}
\epsfig{file=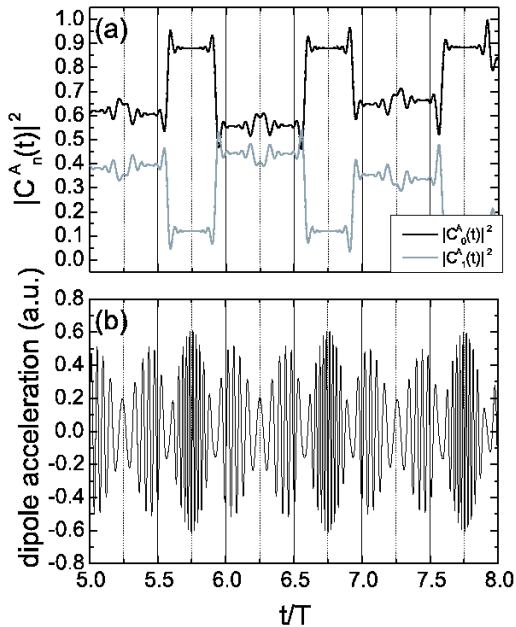,width=7.0cm,angle=0}
\end{center}
\caption{Populations $|C^A_n(t)|^2$ of the adiabatic states (Part (a)) and 
dipole acceleration (Part (b)) as 
functions of time, for a bichromatic field $
E(t)=E_{01}\sin (\omega t)+E_{02}\sin (2\omega t+\theta ),$ with 
$\theta =\pi/2$, 
$\omega=0.05$ a.u., $\omega_{10}=0.409$ a.u., $x_{10}=1.066$
a.u., and field-strength ratio $\zeta =E_{02}/E_{01}=0.8.$ The time $t$ is
given in units of the field cycle.}
\label{bic_4}
\end{figure}

\begin{figure}[tbp]
\begin{center}
\epsfig{file=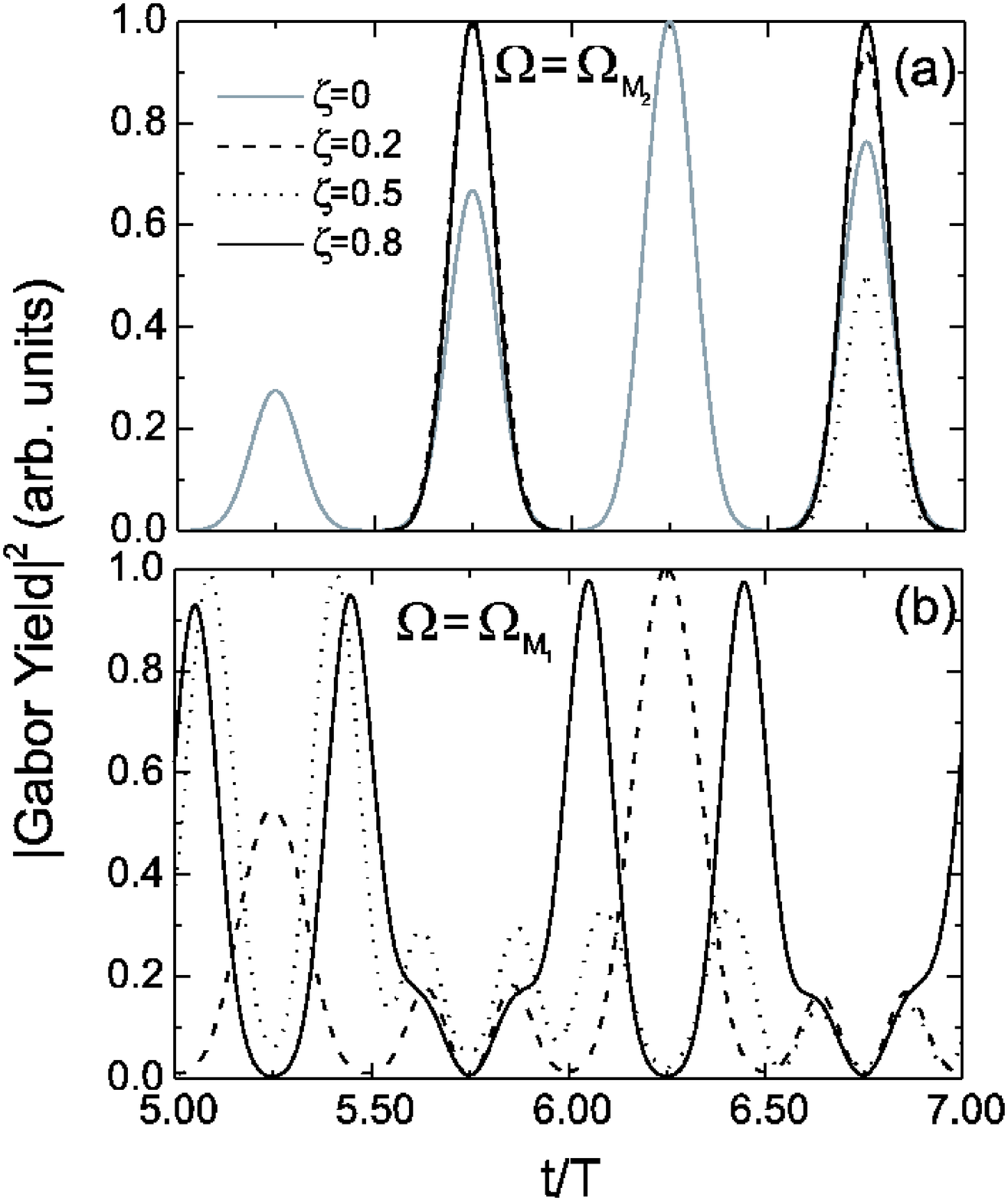,width=7.0cm,angle=0}
\end{center}
\caption{Gabor spectra of the dipole acceleration as a function of time, for
a bichromatic field 
$E(t)=E_{01}\sin (\omega t)$ $+E_{02}\sin (2\omega t+\theta ),$ with 
$\theta =\pi/2$,
$\omega=0.05$ a.u., $\omega_{10}=0.409$ a.u., $x_{10}=1.066$
a.u. and several field-strength ratios $\zeta =E_{02}/E_{01}.$ The maximal
field strength is kept fixed according to Eq. (\protect{\ref{strength2}}) and
equal to $E_{0}=1$ a.u. The upper-cutoff energy lies at
 $\Omega_{M_2}=43\omega$. The lower-cutoff energy varies with $\zeta$. All
cutoff energies are  given in Table III, together with the population transfer times $t_0$ and 
$t_{1M_2}$.
In Part (a), the window function is centered at the upper-cutoff
harmonics ($\Omega_{M_2}=2\varepsilon^A_{M_2}$),
 and the field-strength ratio is $0\leq \zeta \leq 0.8.$ In Part
(b), the center of the window function is taken at
$\Omega_{M_1}=2\varepsilon^A_{M_1}$. All curves have been normalized to
their maximum values. In Part (a), the monochromatic case
is also displayed for comparison.}
\label{gabbic2}
\end{figure}

As in the previous section, we can trace all distortions observed in these
physical quantities back to those observed in time dependence 
of $\varepsilon^A_\pm$. For instance, the shifts in the
level-crossing times $t_0$ predicted by Eq. (\ref{cross2})
 are also present in the main population-transfer times 
for the adiabatic states (Fig. \ref{bic_4}(a)) and in the nodes
of the dipole acceleration (Fig. \ref{bic_4}(b)). Another effect which is
clearly seen in both quantities is the splitting of the maxima near
$t_{1M_1}=0.25T\hspace*{0.1cm} \mathrm{mod}\hspace*{0.1cm} T$. Indeed, there
exist now two sets of maxima which are symmetric with 
respect to these times, for $\zeta\geq 0.25$.
We now investigate the Gabor transform of the
cutoff and plateau harmonics. In Fig. \ref{gabbic2}(a), we display the
time-resolved spectra, centered at the harmonic frequencies
 $\Omega_{M_2}=2\varepsilon^A_{M_2}$,
for different field-strength ratios $\zeta$. The monochromatic case is also
displayed for comparison. As a general feature, for $\zeta\neq 0$, the peaks
of the Gabor spectra at 
$t_{1M_1}=0.25T \hspace*{0.1cm}\mathrm{mod}\hspace*{0.1cm} T$ vanish. This is
a direct consequence of the splitting of the extrema of the adiabatic energies
caused by the high-frequency wave. Due to this splitting, the energy maxima near
$t_{1M_1}$ lie outside the range of the window function and do not contribute
to the time-resolved spectra. Furthermore, as predicted in Eq. (25),
 the peaks at the maxima  
$t_{1M_2}=0.75T \hspace*{0.1cm}\mathrm{mod}\hspace*{0.1cm} T$
 do not move in time as $\zeta$ is varied.

Taking now the window function (\ref{window}) centered at 
 $\Omega_{M_1}=2\varepsilon^A_{M_1}$ (Fig. \ref{gabbic2}(b)), 
one observes, as expected, a completely different behavior for the 
peaks near $t_{1M_1}=0.25T \hspace*{0.1cm}\mathrm{mod}\hspace*{0.1cm} T$. 
For $\zeta< 0.25$, these peaks are exactly at these
times. For $\zeta\geq 0.25$, as expected, they now occur at $t'_{1M}=1/\omega
\arcsin[1/(4\zeta)]$, which  vary with the
field-strength ratio $\zeta$. Furthermore, this second set of peaks splits
for these larger field-strength ratios, such that two 
sets of peaks which are symmetric with respect to 
$t_{1M_1}$ are 
now present. 
Other sets of peaks which can be seen in the picture correspond to the
upper-plateau return times, which occur for $\Omega<\Omega_{M_2}$ and are
symmetric with respect to $t_{1M_2}=0.75T\hspace*{0.1cm}
\mathrm{mod}\hspace*{0.1cm} T$. These peaks  come from the 
splitting of $t_{1M_2}$, which occurs in this energy range (c.f. Fig. \ref{bic_3}).
The population-transfer times for the specific parameters of this figure,
together with the corresponding harmonic frequencies, are
given in Table III.

\subsection{Fourier spectra for the two phases}

In the investigations performed so far, our main objective was to understand
how an additional driving wave may distort the time dependence of the adiabatic
energies and the time profile of harmonic generation. In this section, we address the question of how these distortions 
influence the harmonic spectra. Furthermore, we are interested in extending
the cutoff, and, by doing so, guaranteeing that the
harmonics in this energy region are strong enough for applicational purposes.
Clearly, the ideal scenario is to extend the cutoff energy without any
intensity loss in the corresponding harmonic range. 

\begin{figure}[tbp]
\begin{center}
\epsfig{file=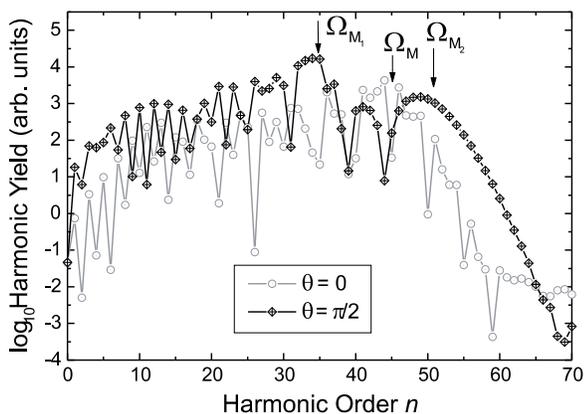,width=8.0cm,angle=0}
\end{center}
\caption{Spectra computed from the dipole acceleration, for the bichromatic
field $
E(t)=E_{01}\sin (\omega t)+E_{02}\sin (2\omega t+\theta ),$ for $\theta=0$,
$\theta=\pi/2$, and field strenghts $E_{01}=1.0$ a.u. and $E_{02}=0.2$ a.u. 
The field is switched on linearly within two cycles.
 The remaining parameters are
$\omega=0.05$ a.u., $\omega_{10}=0.409$ a.u., $x_{10}=1.066$. The cutoff
frequency for $\theta=0$ is roughly at $\Omega_M=46\omega$, whereas for
$\theta=\pi/2$ the cutoff frequencies are approximately at 
$\Omega_{M_1}=35\omega$ and $\Omega_{M_2}=52\omega$. All cutoff energies
are indicated by arrows in the figure.}
\label{comparison}
\end{figure}

With that purpose, we keep $E_{01}$ and $E_{02}$ fixed and compare spectra
obtained for $\theta_1=0$ and $\theta_2=\pi/2$. These results are displayed in
Fig. \ref{comparison}.  As a global 
feature, one
observes that, for $\theta=0$, all harmonics behave in a very similar way, with no distinct
regions, as for instance a double plateau, in the spectra. This is related to
the fact that no splitting of the cutoff energy occurs in this case. The
two maxima in $\varepsilon^A _\pm$ have the same energy, even though the
level-crossing pattern is no longer periodic in $T/2$. On the other hand, for 
$\theta=\pi/2$, there is a clear double-plateau structure. In fact, one 
can identify
a completely different physical behavior for the harmonics in the frequency
regions $\Omega<\Omega_{M_1}$ and 
$\Omega_{M_1}<\Omega<\Omega_{M_2}$. The double-plateau structure is due to the
different cutoff energies which exist in the $\theta =\pi /2$ case. 

 Another generic feature is that the cutoff energy is extended for  
$\theta=\pi/2$. This is
expected, since this quantity is given by the maximum energy difference 
between the adiabatic states.
For a field given by Eq. (\ref{bichr}), the maximal possible energy
is obtained for $E(t_{1M_2})=E_{01}+E_{02}$. This yields the harmonic frequency
 $\Omega_{M_2}$, discussed
in the previous subsection.
 
There exist however non-generic features, which depend on the absolute field
parameters, as for instance its strength. 
Examples of such features are the intensity ratio between the upper and lower
parts of the plateau for $\theta=\pi/2$, and the intensities of the harmonics
obtained for $\theta=\pi/2$, compared to those obtained for
$\theta=0$. Thus, depending on the absolute parameters used, it is not
always possible to extend the cutoff energy without loss of intensity. In
order to control HHG in a two-level atom, a more detailed study of these
features for the particular system in question is necessary.
 These non-generic features are mainly due to the fact that the population 
transfer at the 
level crossings is, in general, given by more complicated expressions than in
the monochromatic case. Indeed, these expressions depend
on the shape and width of the crossing, and on the duration of the
interaction. These shapes have been studied in \cite{SuomG92,shapes}.
 Furthermore, the global structures of the adiabatic-state populations 
$|C^A_n(t)|^2$ have a stronger influence on the spectra in the bichromatic 
case than for monochromatic driving fields.

\section{Scaling behavior}
\label{res3}

In the results discussed in the previous sections, we have used rather
unrealistic frequencies and intensities for the driving fields, for 
which most physical systems would ionize
immediately. This choice of parameters allows us to obtain results with 
 very little numerical effort.  In order to extend our computations 
to more realistic cases, 
as for instance solids, there are two possibilities. Either one slightly
increases the effort to obtain the necessary precision, or one must find 
specific combinations of parameters for which the physical quantities involved 
remain invariant. This second approach has the advantage to provide additional
insight into the physics of the problem.  

With that purpose, we analyze the scaling behavior of these quantities. 
We use scaling laws which have been derived elsewhere 
\cite{FFS2000}, in the context of stabilization of atoms in strong laser 
fields.
 We concentrate on the question of whether driving fields of much lower
frequencies and intensities could originate similar spectra, with, 
for instance, the
same number of harmonics, or the same population-transfer times, 
in units of the field cycle.
Therefore, our starting point will be the expression
\begin{equation}
\sin(\omega t_1)+\zeta \sin(n\omega t_1+\theta)=\pm
\sqrt{\left(N\gamma_1\right)^2-\left(\gamma_2\right)^2}
\label{tgen}
\end{equation}
which relates the harmonic energy to the energy difference of the adiabatic
states. This equation gives the population-transfer times.
 For $\zeta=0$, one has
the monochromatic-field case (Eq. (\ref{tret})), and, for $\zeta\neq 0$ and $n=2$, the
bichromatic situation discussed in the previous section.
 Note that the parameters $E_0$, $\omega$, $\omega_{10}$ and $x_{10}$ 
appear combined, as $\gamma_1=\omega/(2x_{10}E_0)$, or 
$\gamma_2=\omega_{10}/(2x_{10}E_0)$. The denominators of these expressions
give the Rabi frequencies $\Omega_R=2x_{10}E_0$, which scale like the 
energies (c.f. Eqs. (\ref{Hdiab}) and (\ref{Hadiab}) for the two-level 
Hamiltonian). This
keeps the Schr\"odinger Equation invariant under scale transformations.

 We now consider the scale transformation
\begin{equation}
\omega \rightarrow \omega'=\lambda \omega ;\hspace*{0.15cm}
\omega_{10} \rightarrow \omega'_{10}=\lambda \omega_{10}; \hspace*{0.15cm}
\Omega_R \rightarrow \Omega_R'=\lambda \Omega_R,
\end{equation}
where $\lambda$ denotes the dilatation factor. 
The invariance of the Schr\"odinger equation also requires that the time
scales as $t \rightarrow t'=\lambda^{-1} t$,
such that Eq. (\ref{tgen}) will remain invariant.

This apparently trivial result has far-reaching consequences. In fact,
it shows that, for {\it any} set $E_0$, $\omega$, $\omega_{10}$
 and $x_{10}$, the number of harmonics $N$ in the spectra and the 
corresponding population-transfer times $\widetilde{t}_1=\omega t_1/(2\pi)$, 
given in terms of field cycles, remain invariant, as long as $\gamma_1$
and $\gamma_2$ are kept constant. 

Since the unitary transformation (\ref{rotation}) which gives the 
adiabatic states also depends on $E_0$, $\omega$, $\omega_{10}$
 and $x_{10}$ through $\gamma_1$ and $\gamma_2$, it also remains invariant in
this case. Thus, this invariance must also hold for the populations of these 
states, i.e., $|C_{n}^{A}(t)|^2=|C_{n}^A(t')|^2$.

Another quantity of interest is the dipole  acceleration.
 A quick inspection of Eq. (\ref{d2}) shows that 
this quantity does not remain invariant under the
above-stated transformations. In fact, it scales as $x_{10}$ multiplied by the
square of the energy. The dipole
matrix element scales as
 $x_{10} \rightarrow x_{10}'=\lambda^{-1/2} x_{10}$. 
Thus, $\ddot{x}(t)=\lambda^{3/2}\ddot{x}(t')$.

\begin{figure}[tbp]
\begin{center}
\epsfig{file=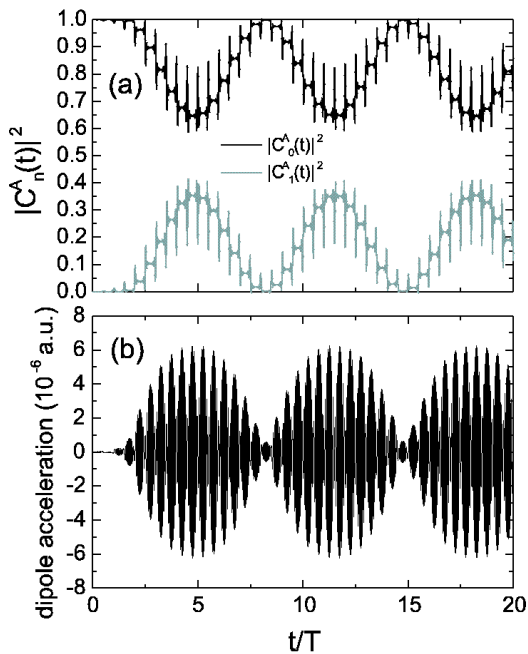,width=7.0cm,angle=0}
\end{center}
\caption{Global structures as functions of time, for: (a) the populations $%
|C_{n}^{A}(t)|^{2}$ of the adiabatic states; (b) the dipole acceleration $%
\ddot{x}(t)$.
The field strength, the field frequency, the transition frequency and the
dipole matrix element were chosen as $E_0=6.71 \times 10^{-6}$ a.u.,
$\omega=2.5 \times 10^{-5}$ a.u., 
 $\omega_{10}=2.045 \times 10^{-5}$ a.u., and $x_{10}=47.673$ a.u., 
respectively.  
These parameters are typical for solid-state systems and give $\gamma_1=0.0391$, $\gamma_2=0.3197$, which are the same
as in Fig. 4. They are obtained from those in Fig. 4
using a scaling transformation with $\lambda=1/2000$. For this set of
parameters, we have used a five times smaller timestep than in the previous 
figures and double precision. The dipole acceleration
is given in atomic units and
the time is given in units of the field cycle. The field is switched on linearly within two cycles. }
\label{scaling1}
\end{figure}

The above-stated conclusions are confirmed by Fig. \ref{scaling1}. 
In this figure, we display the same
physical quantities as in Fig. 4 
for a completely different set of  
parameters which, however, yield the same $\gamma_1$ and $\gamma_2$.
The populations $|C_{n}^{A}(t)|^2$, in this case (c.f. Fig. \ref{scaling1}(a)) are,
as expected, identical to those depicted in Fig. 4. This is true not 
only for the oscillations which are periodic in $T/2$, but also for the 
global enveloping functions. The scaling with $\lambda^{3/2}$ is also observed for
the dipole acceleration (Fig. \ref{scaling1}(b)). The parameters used in the
figure are typical for quantum wells and solid-state systems 
\cite{confine}.

\begin{figure}[tbp]
\begin{center}
\epsfig{file=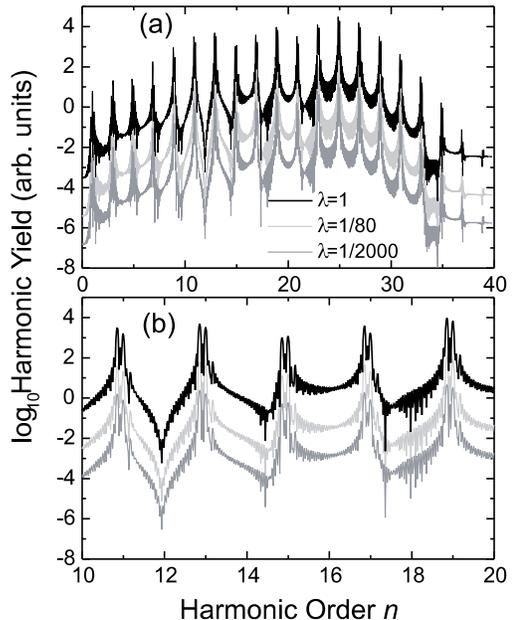,width=7.0cm,angle=0}
\end{center}
\caption{Harmonic spectrum for the same parameters as in Fig. 4 $(\lambda=1)$,
 compared to
those obtained for several field strengths
 $E_0$, field frequencies $\omega$, transition frequencies
 $\omega_{10}$ and matrix dipole elements $x_{10}$, chosen such that 
$\gamma_1=0.0391$ and $\gamma_2=0.3197$, i.e., the same as in Fig. 4.  These
parameters are displayed in Table IV. Part (a)
shows the whole spectra, whereas part (b) displays both spectra for harmonic
order $10<N<20$, such that their substructure can be seen. The field is switched on linearly within two cycles.}
\label{scaling2}
\end{figure}

Another interesting aspect concerns the resulting harmonic spectra. Even
though, in absolute terms, these spectra have different cutoff frequencies and
different global intensities,
for equal $\gamma_1$ and $\gamma_2$ they have the same shape.
Not only the number of harmonics is the same. In addition, all substructure 
in the spectra looks
strikingly similar. These features can be easily understood:
the global intensity decrease is related to the decrease in amplitude of the
dipole acceleration and the identical shapes are a consequence of the fact
that the populations of the adiabatic states, as well as all oscillations
present in the dipole acceleration, remain invariant under the scale
transformations discussed here. 
This is shown in Figs. \ref{scaling2}(a) and (b), for several dilatation
factors $\lambda$. The corresponding field and two-level atom parameters are
given in Table \ref{lbdatable}.

\begin{figure}[tbp]
\begin{center}
\epsfig{file=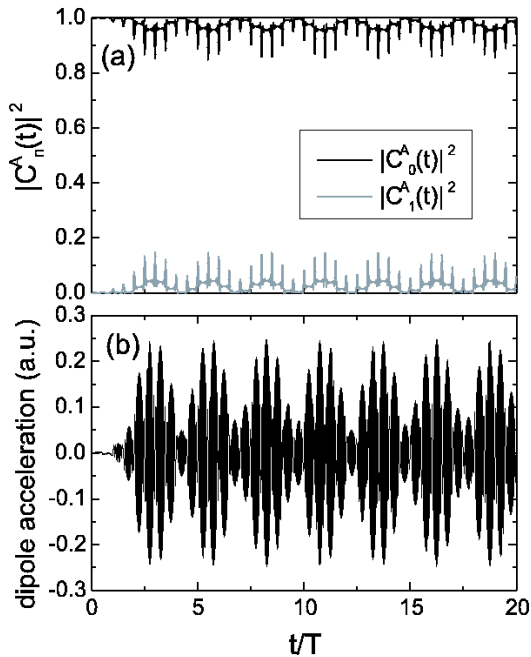,width=7.0cm,angle=0}
\end{center}
\caption{Global structures as functions of time, for: (a) the populations 
$|C_{n}^{A}(t)|^{2}$ of the adiabatic states; (b) the dipole acceleration 
$\ddot{x}(t)$.
The field strength, the field frequency, the transition frequency and the
dipole matrix element were chosen as $E_0=0.62$ a.u., $\omega=0.05$ a.u., 
 $\omega_{10}=0.409$ a.u. and $x_{10}=1.066$ a.u., respectively.  
These parameters are slightly different from the ones in Fig. 4, but give
 $\gamma_1=0.0378$, $\gamma_2=0.3094$. The field is switched on linearly
within two cycles. 
The time is given in units of the field cycle.}
\label{scaling3}
\end{figure}
\begin{figure}[tbp]
\begin{center}
\epsfig{file=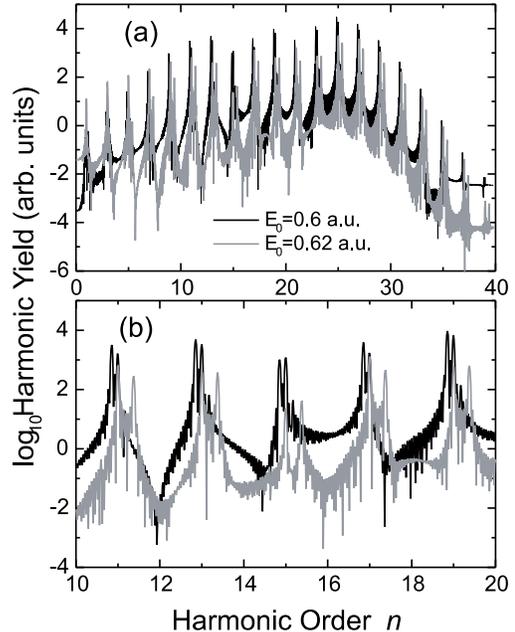,width=7.0cm,angle=0}
\end{center}
\caption{Harmonic spectrum for the same parameters as in Fig. 4, compared to
the one obtained for
 $E_0=0.62$ a.u., $\omega=0.05$ a.u., 
 $\omega_{10}=0.409$ a.u. and $x_{10}=1.066$ a.u., respectively.  
These parameters give  $\gamma_1=0.0378$, $\gamma_2=0.3094$, whereas the ones
 in Fig. 4 yield $\gamma_1=0.0391$, $\gamma_2=0.3197$. Part (a)
shows the whole spectra, whereas part (b) displays both spectra for harmonic
order $10<N<20$, such that their substructure can be seen. The field is switched on linearly within two cycles.}
\label{scaling4}
\end{figure}
On the other hand, the behavior of the system can already be altered by small
variations in  $\gamma_1$ and $\gamma_2$. For instance, in Fig. \ref{scaling3} we
consider a slightly larger field amplitude than in Fig. 4, which give 
different $\gamma_1$ and $\gamma_2$.
In this case, one observes a radically different pattern for the
populations $|C_{n}^{A}(t)|^2$ [Fig. \ref{scaling3}(a)] and the dipole 
acceleration
[Fig. \ref{scaling3}(b)]. As a direct consequence, the spectra does not
exhibit the same substructure [Figs. \ref{scaling4}(a) and (b)].
\section{Conclusions}
\label{concl}

The results discussed in the previous sections lead to the main conclusion
that the three-step model and the two-level atom are not completely different
physical pictures for describing high-harmonic generation, as commonly 
believed.
Indeed, in both models, 
this phenomenon takes place as a result of a
three-step process. Hints that a correspondence between
both physical pictures might exist have
been provided in the literature \cite{gauthey,solid2}. We go however 
beyond such studies, giving evidence
that a three-step mechanism exists in the two-level atom case and
analysing its features in detail. 

In the usual form of the three-step model, there is population transfer from
the atomic ground state to a state in the continuum, i.e., tunneling or
multiphoton ionization. The electron then propagates in the continuum within a
time interval $\tau=t_1-t_0$, gaining a certain amount of kinetic energy which
is converted into harmonic radiation at a time $t_1$, when there is
population transfer from the continuum to the ground state, i.e.,
recombination. 
In the two-level
atom framework, a very similar process takes place: there is population
transfer from the field-dressed state 
$\left| \phi _{0}^{A}(t)\right\rangle$  to the state 
$\left| \phi _{1}^{A}(t)\right\rangle$ at a time $t_0$ for which an avoided
crossing occurs. Subsequently, the 
system acquires energy from the field within the interval $\tau=t_1-t_0$, and,
at a further time $t_1$, when population transfer from
$\left| \phi _{1}^{A}(t)\right\rangle$ back to 
$\left| \phi _{0}^{A}(t)\right\rangle$ takes place, this energy is released 
in form of harmonic radiation. Thus, the main difference between the
three-step model and the two-level atom physical pictures is that in the
latter case, the three steps do not involve a continuum state, but a
field-dressed bound state. 

Further similarities are observed in the time profile of high-harmonic
generation. In both cases, the population transfers which contribute to the
generation of a particular set of harmonics occur at very specific times.
In the usual three-step model, these times are such that the
energy of a particular harmonic must be equal to the
sum of the kinetic energy of the electron upon return and the atomic
ionization potential. The same line of argumentation holds in the 
two-level case, but now the harmonic energy must be equal to the energy
difference between the adiabatic states at these times.

Specifically for monochromatic driving fields, both models share several
features. Both in the three-step model and in the two-level atom case, 
there is a single time corresponding to the generation of the cutoff harmonic.
In the former model, this time corresponds to the maximal kinetic energy the
electron may have, upon return, whereas in the latter model, it gives the
maximal energy difference between the adiabatic states. Also for both cases,
this time splits into two sets of times as the harmonic energy decreases. 
The constructive interference between the corresponding population transfers 
originates the plateau in the high-harmonic spectra.
 This pattern repeats
itself every half cycle of the driving field. This is a direct consequence of
the periodicity of the relevant physical quantities, namely the electron
kinetic energy in the three-step model \cite{footnote} and the adiabatic
energies $\varepsilon^A_\pm$ in the two-level atom case. All these features are
observed as peaks in the Gabor transform of the dipole acceleration. In the
three-step model framework, analogous studies have been performed in \cite{tprof}.

Also for bichromatic driving fields, there are several characteristics 
which are present in both models. A good example is the multiple cutoff 
structure. Indeed, 
the harmonic spectra in this case may exhibit several cutoffs, which,
depending on the model in question, are
given by the maxima of either the electron kinetic energy  or of the
energy difference between the adiabatic states. The number of these cutoffs,
as well as their energies or the corresponding population-transfer times, are
determined by the frequency ratio $n$, the field-strength ratio $\zeta$ and the
relative phase $\theta$. For both the three-step model and the two-level atom,
all peaks in the Gabor spectra
can be traced back to the population-transfer times. In one or the other 
case,  these population transfers occur either between the
adiabatic states (Sec. \ref{res2}), or between the ground-state and the
continuum \cite{cfbichro1}.

Similarities between the two models are also observed for the probability that the ``first step", i.e.,
population transfer, takes
place. In the
three-step model, this probability, per unit time, is roughly given by the
quasi-static tunneling rate ${\cal P}\sim \exp[-{\cal C}/|E(t_0)|]$ \cite{LaLi}. A strong field
$E(t_0)$ at the ionization time $t_0$ yields strong harmonics at the recombination
 time $t_1$.
 This relation is very useful for controlling harmonic spectra, as for
instance the relative intensities of a double plateau (see, e.g., 
\cite{cfbichro1,cfbichro2} for concrete examples).  Within the two-level atom
framework and in the monochromatic case, to first approximation, the
field-dependent terms of the
two-level Hamiltonian can be linearized at the crossings
\cite{gauthey}. Thus, the population transfer between the exchanged states
can be computed by means of the Landau-Zener
model \cite{SuomG92,LandauZener}. This probability is approximately given by 
${\cal P}\sim\exp[-{\cal C}'\pi/(2x_{10}E_0)],$ such that the 
Rabi frequency, in the two-level atom, plays a similar role as $E(t_0)$ in the
three-step model. In general, however, there is not always a simple expression for the 
population transfer at a level crossing \cite{SuomG92,shapes}, 
such that ${\cal P}$ has to be computed
according to the problem at hand. For instance, ${\cal P}$ may be rather complicated
for bichromatic fields. 
This is a limitation for 
controlling high-harmonic spectra in this latter case.

A particularity of the two-level atom is that the very same
distortions caused by the additional field in the field-dressed energies, as
functions of time, are also present in the adiabatic-state populations 
$|C^A_n(t)|^2$ and in the dipole acceleration.
Specifically for the bichromatic field addressed in
this paper, i.e., a $\omega-2\omega$ field, the whole pattern is no longer 
periodic in $T/2$, but in $T$. This is a consequence of the periodicity of the
adiabatic states, which changes with the additional driving wave. A similar
feature occurs in the three-step model framework, due to an analogous change
in the electron kinetic energy upon return (see, e.g., 
\cite{cfbichro1,cfbichro2} for a discussion of this issue).

An interesting issue which is not discussed in this paper concerns the
influence of ionization or feedback mechanisms on the time profiles of
harmonic generation by a two-level atom. In a previous paper it was shown 
that the main
contributions to harmonic generation from a two-level atom whose states 
decayed according to
quasi-static ionization rates  
occurred at minimal field. These
results did not agree with
the bound-bound transitions computed from the numerical solution of the
Schr\"odinger equation for a gaussian potential with two strongly coupled 
bound states
 \cite{cfmdbound}. The strikingly different time profiles obtained in the
present paper for HHG in a closed two-level atom suggest, however, that these
features are stongly influenced by ionization. Therefore, more accurate 
descriptions of ionization and an adequate
feedback mechanism from the continuum would be necessary in the two-level atom
case with unstable levels. The influence of level widths on the population transfer between quantum
states is discussed in \cite{open}.

Finally, there are scaling laws which allow
extending the studies performed in this paper to a broader parameter range.
In fact, we have shown that the important parameters for determining the
physical behavior of the system are $\gamma_1=\omega/(2x_{10}E_0)$, and
$\gamma_2=\omega_{10}/(2x_{10}E_0)$, which denote the ratio of the field and
transition frequencies to the Rabi frequency, respectively. As long as
$\gamma_1$ and $\gamma_2$ are kept constant, driving fields of completely
different strengths and frequencies acting on systems of completely different
energy gaps can yield similar spectra. For bichromatic fields, an additional
requirement for this invariance are
fixed field-strength ratio $\zeta$, field-frequency ratio $n$ and relative
phase $\theta$.

A concrete example of a system for which these properties may be applied is
for instance a quantum well with $\omega_{10} \sim 10^{-4}$ a.u., and $x_{10} \sim 100$ a.u.,
subject to a field of strength $E_0\sim 10^{-5}$ a.u. and
frequency $\omega \sim 10^{-5}$ a.u. \cite{confine}. Transitions between two
subbands in these systems are described very frequently 
by the semiconductor Bloch equations
in the Hartree-Fock approximation \cite{solid3}. In case collective effects
can be neglected, the corresponding
 Hamiltonian reduces to a two-level one-particle Hamiltonian. 
In such a case, the
results of the present paper are expected to be applicable.

\acknowledgements{We thank M. E. Madjet for beneficial discussions,  
A. Fring for useful comments on the manuscript, and
S. W. Kim and T. Chakraborty for providing references.}

\vspace*{3cm}
\begin{table}[tbp]
\caption{Level-crossing times $t_{0}$, population transfer
times $t_{1}$ and the corresponding harmonic energy $\Omega $, for the
parameters of Fig. 2.  The times are
given in units of the period $T=2\pi /\omega $. The harmonic orders,
together with the approximate harmonic energies
in units of the cutoff frequency $\Omega _M$, are given in the remaining
two columns. This pattern repeats itself every half-cycle of the driving 
field.}
\begin{tabular}{cccc}
$t_{0}/T$ & $t_{1}/T$ & harmonic order & $\Omega /\Omega _M$\\ \hline
$0.5$ &$0.25$& 43 & 1\\ \hline
$0.5$ &$0.14$\hspace*{0.5cm}$0.36$& 35 & 0.8\\ 
$0.5$ &$0.09$\hspace*{0.5cm}$0.41$  & 25 & 0.6\\ 
$0.5$ &$0.05$\hspace*{0.5cm}$0.45$ & 17 & 0.4\\ 
\end{tabular}
\end{table}

\onecolumn
\begin{table}[tbp]
\caption{ Times for the population transfers between the extrema of
the adiabatic
states, with the approximate order of the corresponding cutoff harmonic, for a
bichromatic field given by Eq. (\protect{\ref{bichr}}),
with relative phase $\theta =0$ and several field-srength ratios $\zeta
=E_{02}/E_{01}.$ The field and two-level atom parameters are the same as those used
in Fig \protect{\ref{gabbic1}}. No entry means that corresponding maxima do not exist. 
This pattern repeats itself every cycle $T=2\pi/\omega$ of the
driving field. }
\label{times_1}
\begin{tabular}{ccccccccc}
\multicolumn{3}{c}{$\zeta =0.2$} & \multicolumn{3}{c}{$\zeta =0.5$} & 
\multicolumn{3}{c}{$\zeta =0.8$} \\ 
$t_0/T$ & $t_{1M}/T$ & $\Omega_{M}/\omega$ & $t_0/T$ & $t_{1M}/T$
& $\Omega_{M} /\omega$ & $t_0/T$ & $t_{1M}/T$ & $%
\Omega_{M}/\omega$ \\ \hline
\vspace*{-0.5ex} &  &  &  &  &  &  &  &  \\ 
0 &0.20 & 43& 0 &0.17  & 43 & 0 & 0.15 & 43 \\ 
0.5 &0.80 & 43 &0.5  & 0.83 & 43 & 0.36 & 0.42 & 9 \\ 
 - & - & - & - & - & - & 0.5 & 0.85 & 43\\ 
- & - & - & - & - & - & 0.64 & 0.58 & 9 \\ 
\end{tabular}
\end{table}
\begin{table}[tbp]
\caption{ Times for the population transfers between the extrema of
the adiabatic
states, with the approximate order of the corresponding cutoff harmonic, for a
bichromatic field given by Eq. (\protect{\ref{bichr}}),
with relative phase $\theta =\pi /2$ and several field-srength ratios $\zeta
=E_{02}/E_{01}.$ The field and two-level atom parameters are the same as those used
in Fig \protect{\ref{gabbic2}}. No entry means that corresponding maxima do not exist. 
This pattern repeats itself every cycle $T=2\pi/\omega$ of the
driving field. For $\zeta=0.8$, there are additional avoided crossings
 at $0.25T \mathrm{mod}\hspace*{0.1cm} T$.}
\label{times_2}
\begin{tabular}{ccccccccc}
\multicolumn{3}{c}{$\zeta =0.2$} & \multicolumn{3}{c}{$\zeta =0.5$} & 
\multicolumn{3}{c}{$\zeta =0.8$} \\ 
$t_0/T$ & $t_{1M}/T$ & $\Omega_{M}/\omega$ & $t_0/T$ & $t_{1M}/T$
& $\Omega_{M} /\omega$ & $t_0/T$ & $t_{1M}/T$ & $%
\Omega_{M}/\omega$ \\ \hline
\vspace*{-0.5ex} &  &  &  &  &  &  &  &  \\ 
0.53& 0.75 & 43 & 0.56 & 0.75 & 43 & 0.58 &0.75 & 43 \\ 
0.97&1.25 & 30 & 0.94 & 1.08 & 23 & 0.92 & 1.05 & 24 \\ 
- & - & - & 0.94 & 1.42 & 23 & 0.92 & 1.45 & 24 \\ 
- & - & - & - & - &- & 1.25 & 1.45 & 24 
\end{tabular}
\end{table}
\twocolumn
\begin{table}[tbp]
\caption{Field and two-level atom parameters, given in atomic units, together
with the dilatation factor $\lambda$. All parameters have been chosen such 
that $\gamma_1=0.0391$ and $\gamma_2=0.3197$.}
\begin{tabular}{ccccc}
$x_{10}$ & $E_{0}$ & $\omega$ & $\omega_{10}$&$\lambda$\\ \hline
$1.066$ &$0.6$& $0.05$ & $0.409$ &$1$\\ \hline
$9.535$ &$8.385\times10^{-4}$& $6.25\times10^{-4}$&$5.1125\times10^{-3}$ & $1/80$\\ 
$47.673$ &$6.71\times10^{-6}$ & $2.5\times10^{-5}$ & $2.045\times10^{-4}$ & $1/2000$\\ 
\end{tabular}
\label{lbdatable}
\end{table}


\begin{references}
\bibitem{footnote} The kinetic energy in this case is given by 
$E_{\mathrm kin}(t_1,t_0)=1/2[A(t_1)-A(t_0)]^2$, where $A(t)$ is the vector
potential. Clearly, if $A(t+\pi/\omega)=\pm A(t)$, which is the case for
monochromatic and bichromatic fields with odd frequency ratio, 
$E_{\mathrm kin}(t_0,t_1)=E_{\mathrm kin}(t_0+\pi/\omega,t_1+\pi/\omega)$.
\end{references}
\end{document}